\documentclass[letterpaper,twocolumn,prl,showpacs]{revtex4}
\usepackage[latin9]{inputenc}
\usepackage{amsmath}
\usepackage{amssymb}
\usepackage{graphicx}
\usepackage{esint}

\makeatletter

\@ifundefined{textcolor}{}
{%
 \definecolor{BLACK}{gray}{0}
 \definecolor{WHITE}{gray}{1}
 \definecolor{RED}{rgb}{1,0,0}
 \definecolor{GREEN}{rgb}{0,1,0}
 \definecolor{BLUE}{rgb}{0,0,1}
 \definecolor{CYAN}{cmyk}{1,0,0,0}
 \definecolor{MAGENTA}{cmyk}{0,1,0,0}
 \definecolor{YELLOW}{cmyk}{0,0,1,0}
}

\usepackage{epsfig}\usepackage{color}\usepackage{booktabs}

\makeatother

\begin{document}

\title{Temporal and spatial heterogeneity in aging colloids: a mesoscopic
model }

\author{Nikolaj Becker and Paolo Sibani}

\affiliation{FKF, Syddansk Universitet, DK5230 Odense M, Denmark}

\author{Stefan Boettcher and Skanda Vivek}

\affiliation{Department of Physics, Emory University, Atlanta, GA, USA}

\begin{abstract}
A model of dense hard sphere colloids building on simple notions of
particle mobility and spatial coherence is presented and shown to
reproduce results of experiments and simulations for key quantities
such as the intermediate scattering function, the particle mean-square
displacement and the $\chi_{4}$ mobility correlation function. All
results are explained by two emerging and interrelated dynamical properties:
\emph{i)} a rate of intermittent events, \emph{quakes}, which decreases
as the inverse of the system age $t$, leading to $\mu_{{\rm q}}(t_{{\rm w}},t)\propto\log(t/t_{{\rm w}})$
as the average number of quakes occurring between the `waiting time'
$t_{{\rm w}}$ and the current time $t$; \emph{ii)} a length scale
characterizing correlated domains, which increases linearly in $\log t$.
This leads to simple and accurate scaling forms expressed in terms
of the single variable, $t/t_{{\rm w}}$, preferable to the established
use of $t_{{\rm w}}$ and of the lag time $\tau=t-t_{{\rm w}}$ as
variables in two-point correlation functions. Finally, we propose
to use $\chi_{4}\left(t_{{\rm w}},t\right)$ experimentally to extract
the growing length scale of an aging colloid and suggest that a suitable
scaling of the probability density function of particle displacement
can experimentally reveal the rate of quakes. 
\end{abstract}
\maketitle
Aging is a spontaneous off-equilibrium relaxation process, which entails
a slow change of thermodynamic averages. In amorphous materials with
quenched disorder~\cite{Struik78,Nordblad86,Rieger93,Kob00,Crisanti04,Sibani05},
measurable quantities such as the thermo-remanent magnetization~\cite{Kenning06}
and the thermal energy~\cite{Crisanti04,Sibani07,Sibani08,Christiansen08}
decrease, on average, at a decelerating rate during the aging process.
In dense colloidal suspensions, light scattering~\cite{Cipelletti00,Elmasri05}
and particle tracking techniques~\cite{Weeks00,Courtland03,Lynch08,Candelier09}
have uncovered intermittent dynamics and a gradual slowing down of
the rate at which particles move. Intermittency suggest a hierarchical
dynamics, instead of coarsening, as the origin of this process. However,
changes in spatially averaged quantities such as energy and particle
density are difficult to measure and the question of which physical
properties are actually evolving in an aging colloid~\cite{PhysRevE.75.050404,Cianci06}
lacks a definite answer.

In a recent paper~\cite{BoSi09}, two of the authors proposed that
kinetic constraints bind colloidal particles together in `clusters'.
As long as a cluster persists, its center of mass position remains
fixed, on average, but once it breaks down the particles which belong
to it can \emph{move} independently in space, and are able to \emph{join}
other clusters. The dynamics is controlled by the probability per
unit of time, $P(h)$, that a cluster of size $h$ collapses through
a \emph{quake}. Specifically, if the cluster-collapse probability
is exponential, as in Eq.~(\ref{eq:Ph}) below, quakes follow a Poisson
process whose average is proportional to the logarithm of time. Log-Poisson
processes describe the aging phenomenology of a wide class of glassy
systems~\cite{Crisanti04,Sibani07,Sibani08,Christiansen08,Kenning06}
and, specifically in our case, imply that particle motion is (nearly)
diffusive on a logarithmic time scale, as found in our analysis~\cite{BoSi09}
of tracking experiments~\cite{Courtland03}.

In this Letter, we introduce a model of cluster dynamics based on
these principles that explicitly accounts for the \emph{spatial} form
of the clusters on a lattice in any dimension. This enables us to
calculate the internal energy, the intermediate scattering function,
the mean square displacement, the mobility correlation function $\chi_{4}$
and other measures of spatial complexity. We can hence compare with
simulational~\cite{ElMasri10} and experimental \cite{Berthier05,Berthier11}
data, and suggest a number of experiments to probe colloidal dynamics.

\begin{figure}
\includegraphics[width=0.22\textwidth]{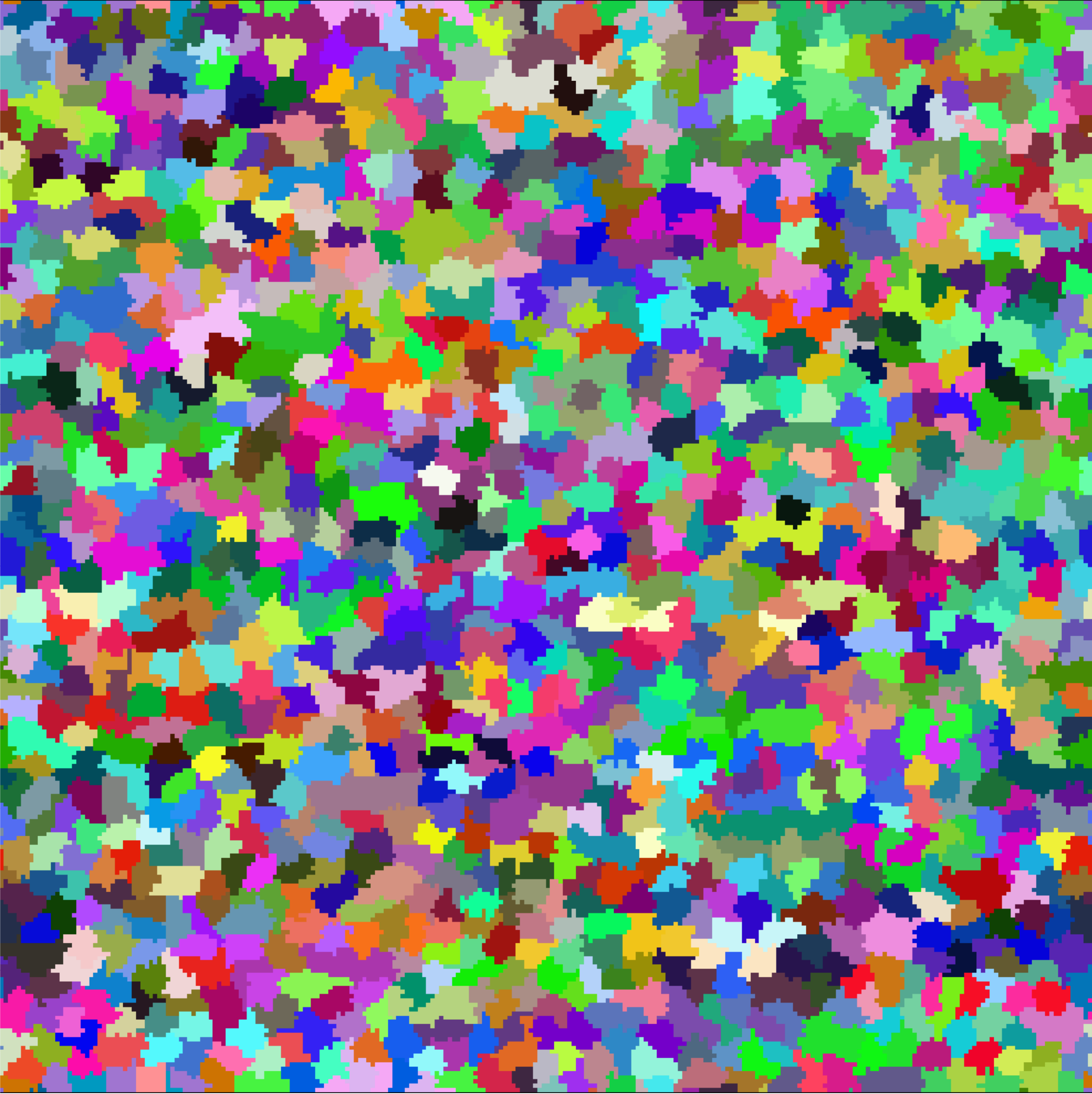}\includegraphics[scale=0.16]{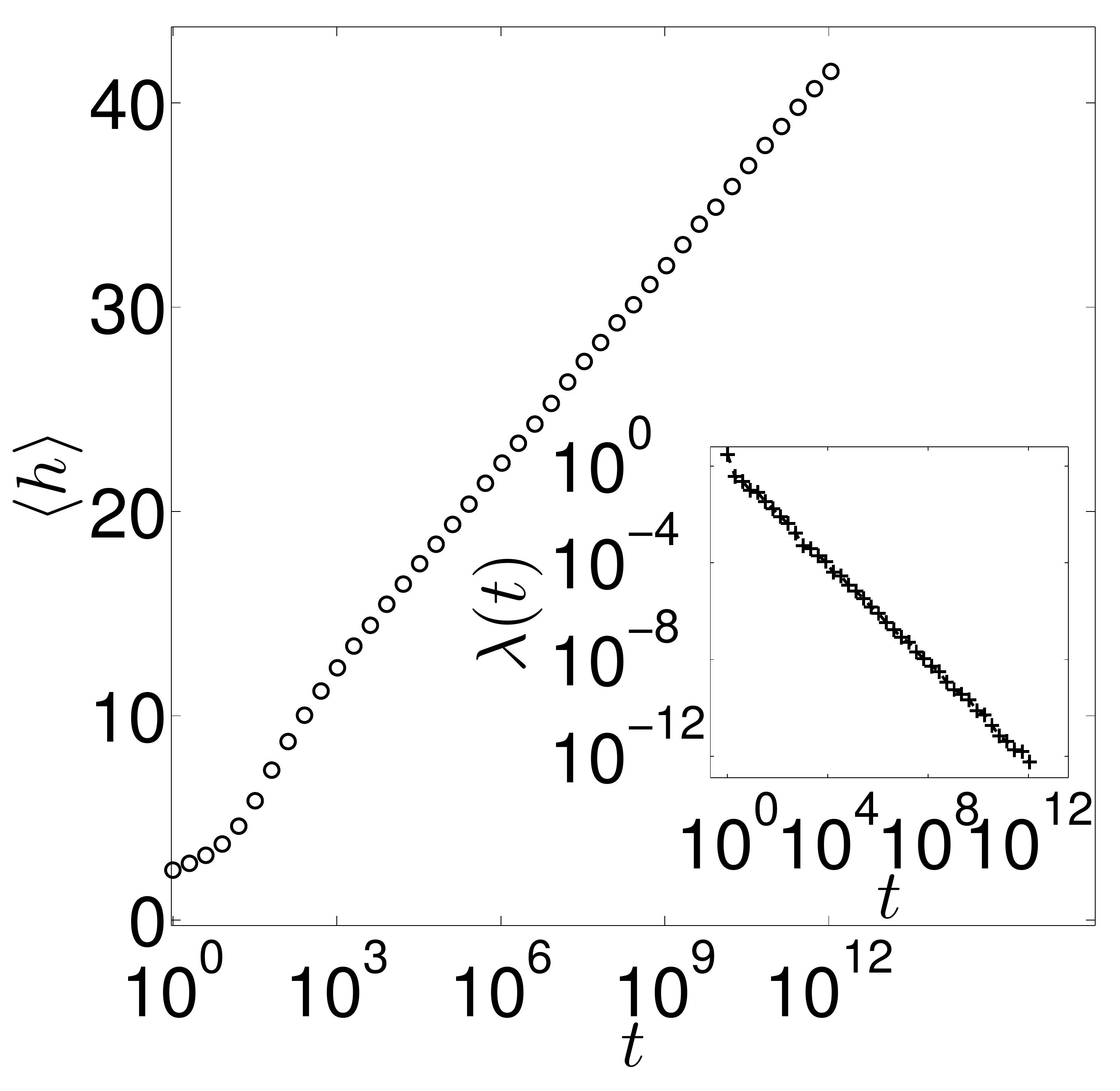}
\vspace{-0.25cm}
 \caption{(a) Snapshot of a $256\times256$ system after $t=10^{15}$ sweeps.
Random colors are assigned to different clusters for visibility. Cluster
sizes apparently are peaked around some average value $\langle h\rangle\sim\log t$,
as shown in (b). The decelerating rate $\lambda(t)\sim1/t$ of cluster
break-up events that emerges from the probability in Eq.~\eqref{eq:Ph}
is shown in the inset.}

\vspace{-0.25cm}
 \label{fig:cluster_map_L256_t50_w1024} 
\end{figure}

Our particles reside on a lattice with periodic boundary conditions,
each lattice site occupied by exactly one particle. Particles are
either mobile singletons (cluster-size $h=1$) or form immobile contiguous
clusters of size $h>1$. When picked for an update, mobile particles
exchange position with a randomly selected neighbor and join that
neighbor's cluster. If the particle is not mobile, either its entire
cluster ``shatters'' into $h$ newly mobile particles with probability
\begin{equation}
P\left(h\right)=e^{-h},\label{eq:Ph}
\end{equation}
or no action is taken. When starting with an initial state consisting
of singletons, i.e., without any structure, the model develops spatially
heterogeneous clusters with a length scale growing logarithmically
in time. The state of the system on a square lattice with $L=256$
after $10^{15}$ sweeps is depicted in Fig.~\ref{fig:cluster_map_L256_t50_w1024}(a)
while Fig.~\ref{fig:cluster_map_L256_t50_w1024}(b) shows the logarithmic
growth of the average cluster size.

As previously shown \cite{BoSi09}, the rate of events decelerates
as $1/t$, see inset of Fig.~\ref{fig:cluster_map_L256_t50_w1024}(b),
which makes random-sequential updates inefficient. In our simulations,
we therefore use the \emph{Waiting Time Method} \cite{Bortz75,Dall01},
where a random ``lifetime'' is assigned to each cluster based on
the geometric distribution associated with $P(h)$; the cluster with
the shortest remaining lifetime is shattered and lifetimes for other
pre-existing or newly formed clusters are adjusted or newly assigned,
following the Poisson statistics. With this event-driven algorithm,
we have been able to follow our model evolution over 15 decades in
time, far exceeding current experimental time windows.

Important aspects of aging dynamics are described by observable quantities
with two time arguments. Here, we denote by $t$ the current time
and by $t_{{\rm w}}$ the \emph{waiting time} before measurements
are taken for a system initialized at $t=0$. To conform to common
usage, the lag time $\tau\equiv t-t_{{\rm w}}$ is used as abscissa
in the main plot of relevant figures. However, we also provide a collapse
of the data, which is best accomplished when \emph{global} time $t$
is scaled by $t_{{\rm w}}$ as independent time variable.

\begin{figure}
\vspace{-0.25cm}
 \includegraphics[width=0.5\textwidth]{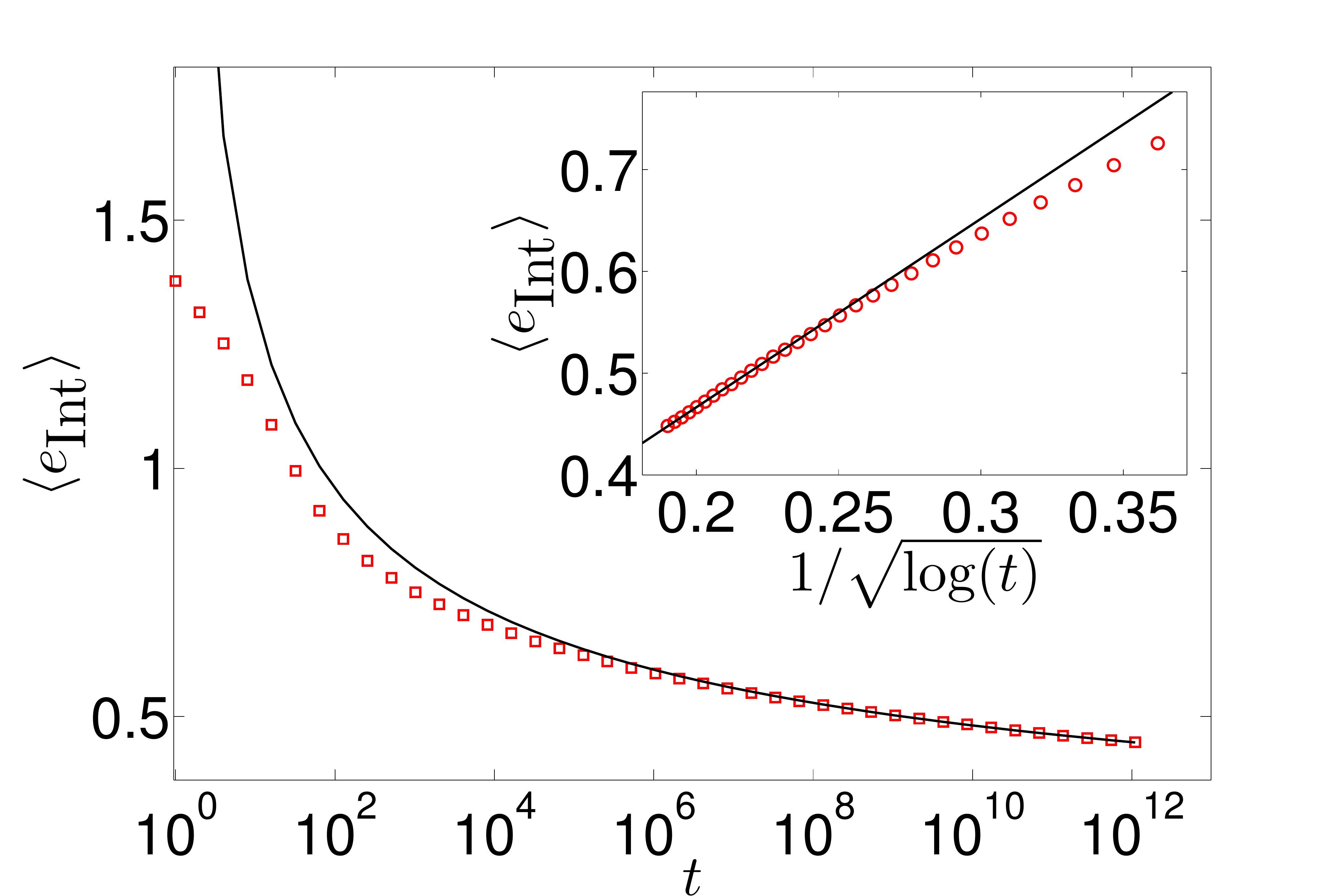}
\caption{Average interface energy per particle, $\left\langle e_{{\rm Int}}\right\rangle $,
for $L=64$. For large $t$, the interface energy follows the form
$\left\langle e_{{\rm Int}}\right\rangle \sim1/\sqrt{\log(t)}$ derived
in Eq. (\ref{eq:ei}), as confirmed by the inset.}

\vspace{-0.5cm}
 \label{fig:mie} 
\end{figure}

Using the details of the Lennard-Jones potential in their molecular
dynamics simulation, El-Masri et al. \cite{ElMasri10} were able to
determine the evolution of the internal energy of a colloidal system
in terms of its pressure. We simply monitor the interface between
clusters as a proxy of the internal energy, assuming that a shrinking
interface indicates a decline in free volume which allows particles
within clusters to relieve their mutual repulsion. The average number
of clusters $\langle n\rangle$ can be written in terms of average
cluster size $\langle h\rangle$ as $\langle n\rangle=L^{2}/\langle h\rangle$.
Since the average cluster size increases with $\langle h\rangle\sim\log(t)$,
see Fig.~\ref{fig:cluster_map_L256_t50_w1024}(b), and since for
compact clusters in two dimensions the interface-length scales as
$S(h)\propto\sqrt{\langle h\rangle}$, the average energy per particle
$\left\langle e_{{\rm Int}}\right\rangle $ is estimated as 
\begin{equation}
\left\langle e_{{\rm Int}}\right\rangle =S(h)\frac{\langle n\rangle}{L^{2}}\sim\frac{1}{\sqrt{\langle h\rangle}}\sim\frac{1}{\sqrt{\log t}}.\label{eq:ei}
\end{equation}
Fig.~\ref{fig:mie} shows that the approximation holds after a more
rapid initial decay. The slow decay matches that of the Lennard-Jones
simulations in Fig.~1 of Ref.~\cite{ElMasri10}, and it is reminiscent
of granular compactification~\cite{BenNaim98}, where noisy tapping
slowly anneals away excess free volume. The same process drives our
cluster growth, although density changes are not explicitly expressed
in the model.

Readily available through light scattering experiments, the \emph{self-intermediate
scattering function} (SISF) $f_{s}$ assesses two-time correlations
used to resolve dynamical characteristics of non-equilibrium systems.
Formally, it is defined as the spatial Fourier transform, 
\begin{equation}
f_{s}\left(\vec{\textbf{q}},t_{{\rm w}},t\right)=\int d\vec{\textbf{r}}\,\mathcal{G}_{s}\left(\vec{\textbf{r}},t_{{\rm w}},t\right)\exp\left[-i\vec{\textbf{q}}\cdot\vec{\textbf{r}}\right],\label{eq:f_s}
\end{equation}
of the self-part of the van Hove distribution function, 
\begin{equation}
\mathcal{G}_{s}\left(\vec{\textbf{r}},t_{{\rm w}},t\right)=\frac{1}{N}\sum_{j}\delta\left[\vec{\textbf{r}}-\Delta\vec{\textbf{r}}_{j}\right],\label{eq:VHD}
\end{equation}
with $\Delta\vec{\textbf{r}}_{j}\left(t_{{\rm w}},t\right)=\vec{\textbf{r}}_{j}\left(t\right)-\vec{\textbf{r}}_{j}\left(t_{{\rm w}}\right)$
as displacement of particles $j$ in the time interval between $t_{{\rm w}}$
and $t$. In general, SISF can be interpreted as a measure of the
``reciprocal of movement'', meaning the average tendency of particles
to stay confined in cages whose size scales with inverse magnitude
of the wave vector $\vec{\textbf{q}}$. Using symmetry and the integer
values of the positions, the discrete version of SISF reduces to 
\begin{equation}
f_{s}(q,t_{{\rm w}},t)=\left\langle \frac{1}{N}\sum_{j=1}^{N}\cos\left(\vec{\textbf{q}}\cdot\Delta\vec{\textbf{r}}_{j}\right)\right\rangle .\label{eq:fqttw}
\end{equation}
Due to spatial isotropy, the SISF is only a function of the magnitude
$q$, with $q_{{\rm min}}=2\pi/L\le q\le\pi\sqrt{2}=q_{{\rm max}}$.

\begin{figure}
\includegraphics[width=0.5\textwidth]{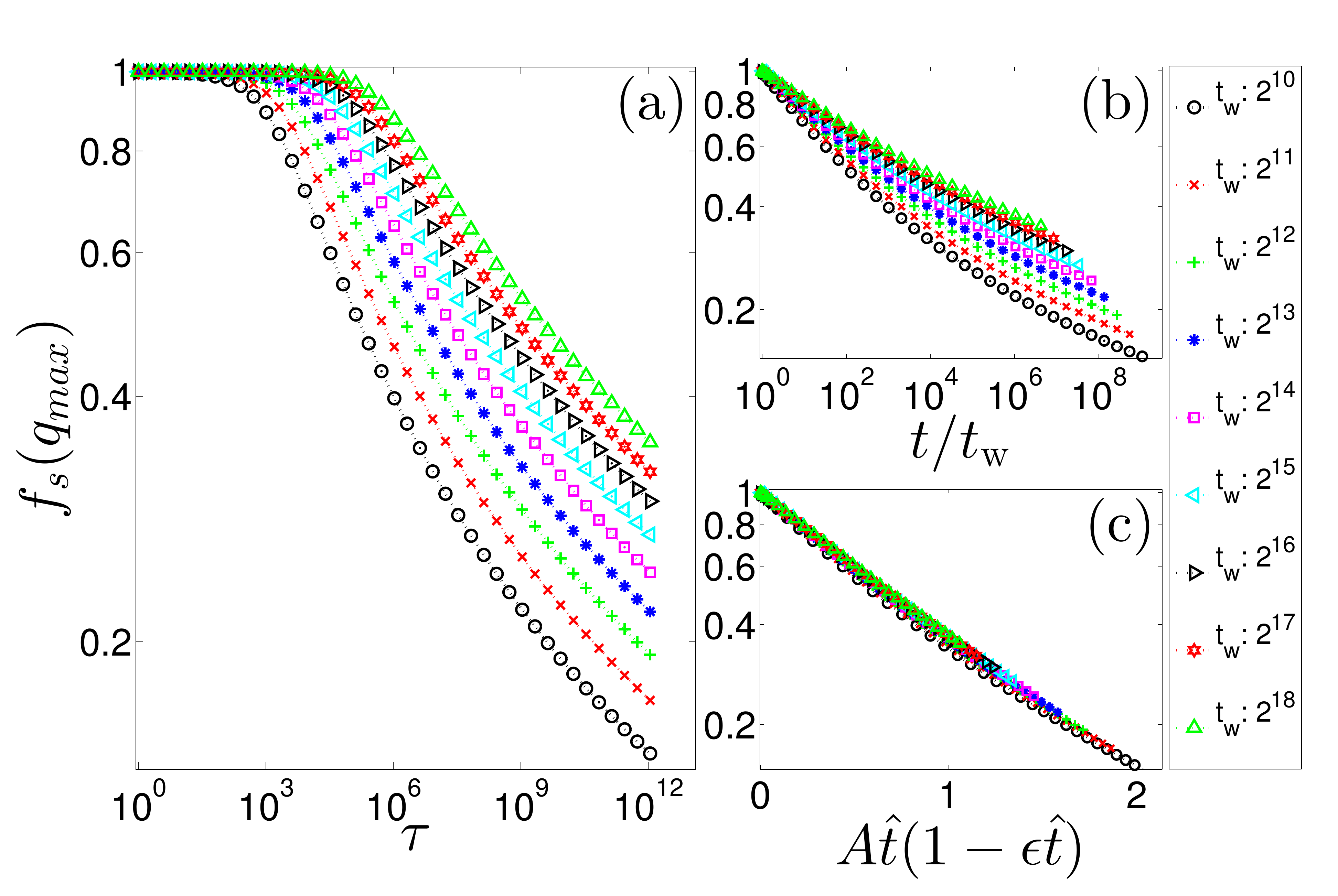}
\vspace{-2.5mm}
 \caption{Decay of SISF at $q_{{\rm max}}=\sqrt{2}\pi$ and system size $L=64$
for $t_{{\rm w}}=2^{k}$, $k=10,11,\ldots,18$, using three different
forms of independent variable: $(a)$ the lag time $\tau$, $(b)$
the total run time $t$ scaled by waiting time $t_{{\rm w}}$, and
$(c)$ the correction to $\hat{t}=\log(t/t_{{\rm w}})$ in Eq.~\eqref{eq:f_s_approx}
(see text for details). The fitted values $A$ and $\epsilon$ remain
approximately constant across the range of different $t_{{\rm w}}$;
with $A\approx0.1$ and a correction of merely $\epsilon\approx1$\%.}

\label{fig:sisf} 
\end{figure}

Figure \ref{fig:sisf} shows the results of simulating an $L=64$
system using $2000$ instances and waiting times varying from $2^{10}$
to $2^{18}$ in powers of two. Panel $(a)$ depicts the behavior of
SISF as a function of lag time. For large $t_{{\rm w}}$, to a good
approximation the data can be represented by a power law of the scaling
variable $\hat{t}=\log\left(t/t_{{\rm w}}\right)$,
$f_{s}\sim C\hat{t}^{-A}$, where 
 $C$ is a constant and $A$ a  positive, non-universal exponent,
see panel $(b)$ of Fig.~\ref{fig:sisf}. Some curvature remains,
nevertheless, and data do not completely collapse. In contrast, panel
$(c)$ achieves excellent  scaling and collapse using the form (discussed
later) 
\begin{equation}
f_{s}\propto\exp\big\{-A\hat{t}\left(1-\epsilon\hat{t}\right)\big\},\label{eq:f_s_approx}
\end{equation}
with $\epsilon\approx0.012$. The power-law exponent $A\approx0.1$
weakly depends on $t_{{\rm w}}$ and changes systematically by about
$40\%$ over two decades of $t_{{\rm w}}$, likely reflecting the
effect of higher order corrections in Eq.~\eqref{eq:f_s_approx}.
Note that $A$ is comparable to the same exponent found in expensive
Lennard-Jones simulations, see Fig.~2 of Ref.~\cite{ElMasri10}.

\begin{figure}
\vspace{-5mm}
 \centering \includegraphics[angle=90,width=0.5\textwidth,height=7cm,keepaspectratio]{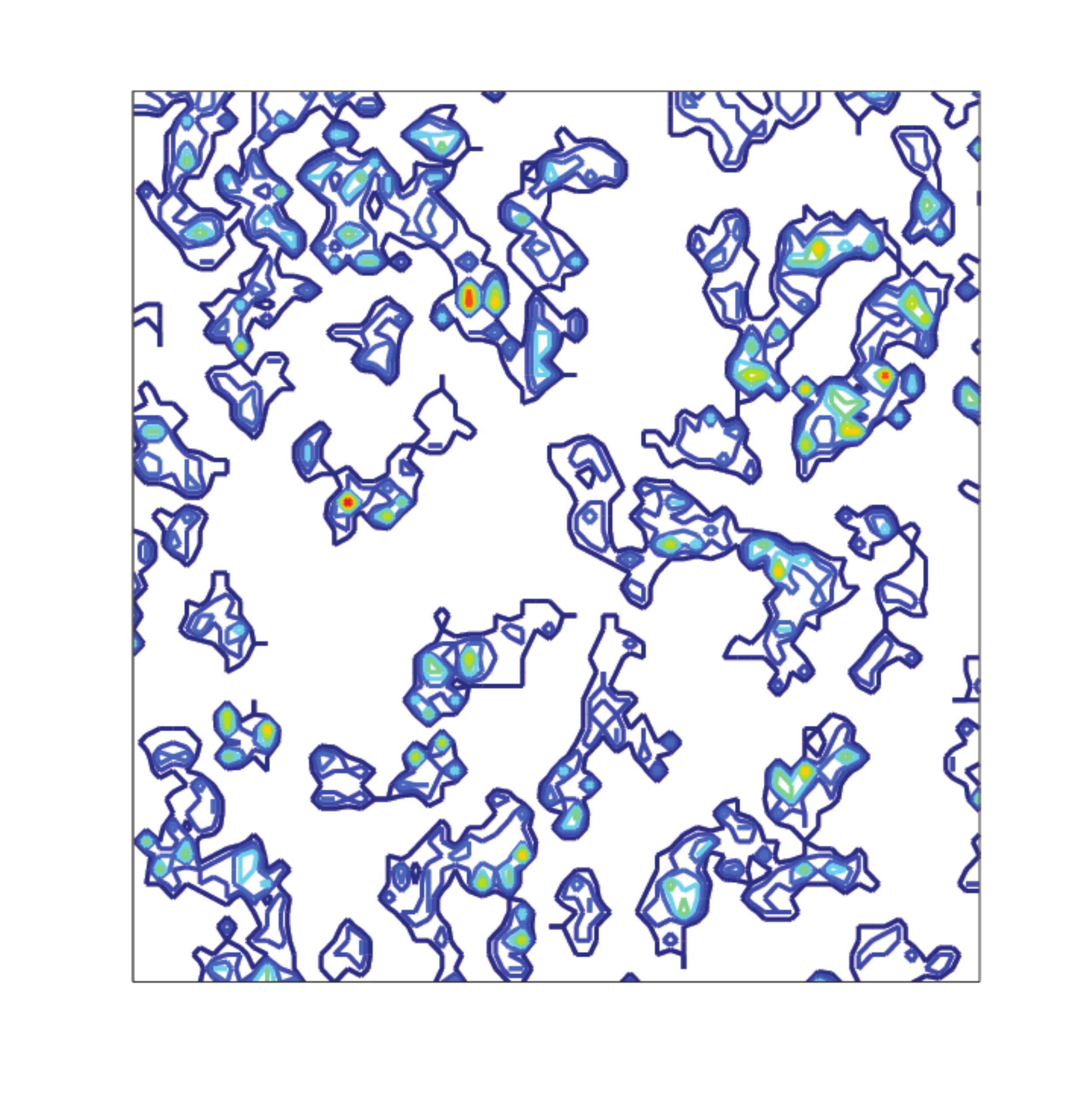}
\vspace{-10mm}
 \caption{Persistence on a $64\times64$ lattice. For times $t$ from $t_{{\rm w}}=2^{12}$
to $4t_{{\rm w}}$ activations are recorded. Within a dominant inert
background (white), domains are encircled with darker to lighter contours
to mark one to ten activations. This mobility pattern demonstrates
dynamic heterogeneity and the associated the preferential return of  activity  to
the same  sites.}

\label{fig:Snapshot} \vspace{0cm}
 
\end{figure}

An alternative characterization of immobility, \emph{persistence},
measures the average fraction of particles that never move \cite{BoSi09,Pastore11}.
Conceptually simple and easily accessible in simulations, persistence
must be deduced indirectly in experiments from the SISF at its peak
wave-vector. In terms of Eq.~\eqref{eq:VHD}, persistence is defined
as 
\begin{equation}
\mathcal{P}\left(t_{{\rm w}},t\right)=\mathcal{G}_{s}\left(\left|\vec{\textbf{r}}\right|\leq d,t_{{\rm w}},t\right),\label{eq:persistence}
\end{equation}
i.e., the fraction of particles that have coordinates at times $t_{{\rm w}}$
and $t$ with $\lvert\vec{\textbf{r}}(t)-\vec{\textbf{r}}(t_{{\rm w}})\rvert\leq d$,
where $d$ is a threshold representing the largest distance a particle
can move without detection. For small $d$, the SISF in Eq.~\eqref{eq:fqttw}
reduces to persistence for $\lvert\vec{\textbf{q}}\rvert\rightarrow\infty$.
To avoid over-counting particles that return to their original position,
in simulations we only count particles that have been \emph{activated}
since $t_{{\rm w}}$. Our results for persistence with $d=1$ are
virtually indistinguishable from those for SISF at $q_{{\rm max}}$
in Fig.~\ref{fig:sisf}. Figure \ref{fig:Snapshot} illustrates the
recurrance of quake activity to sites on a $L=64$ lattice for a time
interval $[t_{{\rm w}},4t_{{\rm w}}]$ with $t_{{\rm w}}=2^{12}$.
While most particles persist in their position, mobility concentrates
in areas scattered about the system (``dynamic heterogeneities''\cite{Weeks00})
with future activity favoring previously mobile sites. 

\begin{figure}[bh]
\vspace{-5mm}
\centering \includegraphics[width=0.5\textwidth]{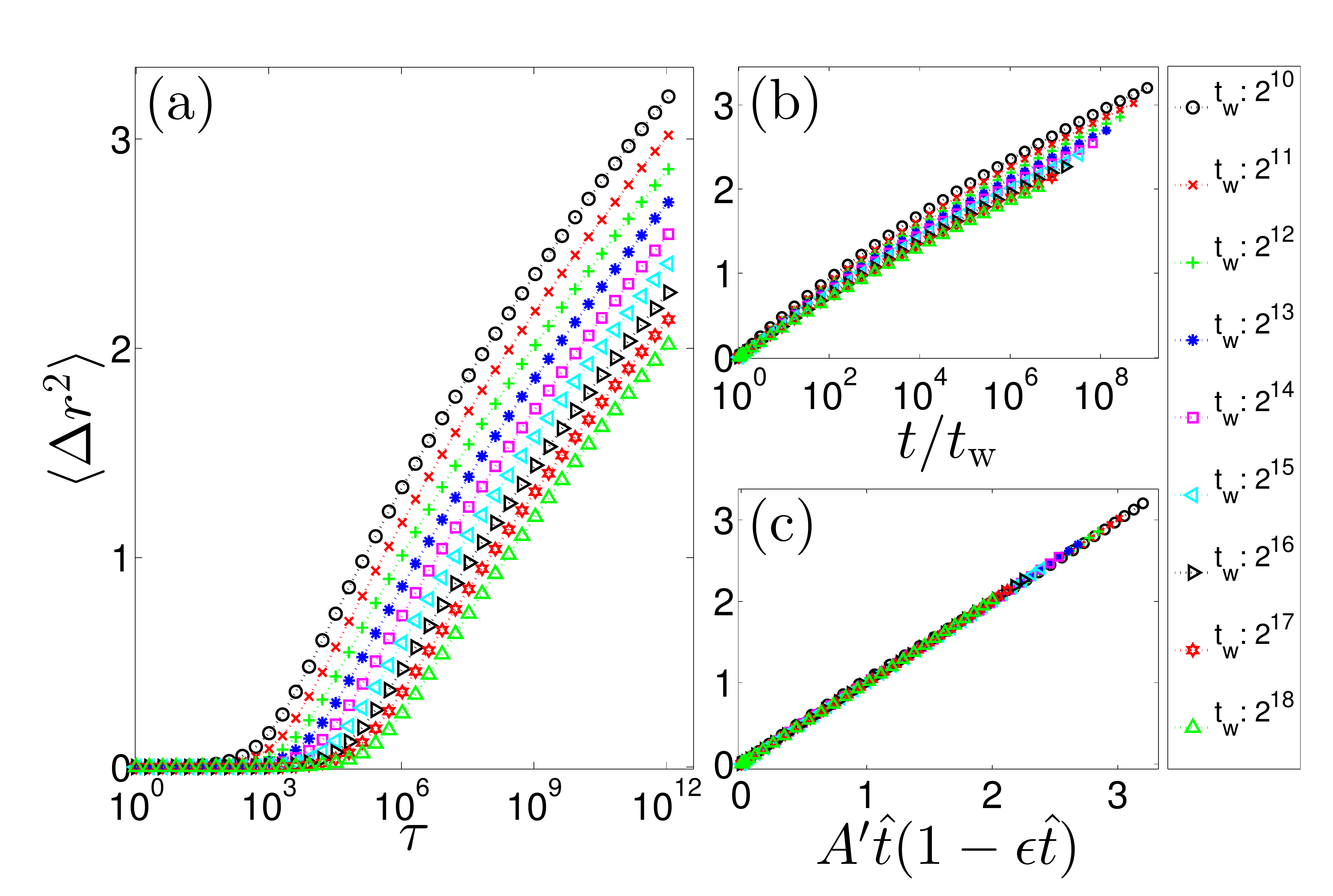}
\vspace{-3.5mm}
 \caption{MSD for a range of different waiting times and different choices of
independent variable. The system parameters are the same as in Fig.~\ref{fig:sisf}. }

\label{fig:msd} \vspace{-0.35cm}
 
\end{figure}

The positional variance or mean square displacement (MSD) between
times $t_{{\rm w}}$ and $t$ is computed by averaging the square
displacement, first over of all particles and then over the ensemble.
Using $\left\langle \cdot\right\rangle $ for the ensemble average
and $\lvert\cdot\rvert$ for the Euclidean norm, the MSD  is written as
\begin{equation}
\Delta r^{2}(t_{{\rm w}},t)=\left\langle \frac{1}{N}\sum_{j=1}^{N}\lvert\vec{\textbf{r}}_{j}(t)-\vec{\textbf{r}}_{j}(t_{{\rm w}})\rvert^{2}\right\rangle .\ \label{eq:msd}
\end{equation}
Figure \ref{fig:msd} shows the MSD for a system of size $L=64$ with
waiting times $t_{{\rm w}}=2^{k}$ for $k=10,11\ldots18$. In analogy
with Fig.~\ref{fig:sisf}, the MSD is plotted vs.~three different
variables: Panel $(a)$ uses the lag time, panel $(b)$ the scaling
variable $t/t_{{\rm w}}$, and panel $(c)$ uses the same type of
correction as in Eq.~\eqref{eq:f_s_approx}, that is, $\Delta r^{2}\propto A'\hat{t}\left(1-\epsilon\hat{t}\right)$
with the same $\epsilon=0.012$ and the ``log-diffusion'' constant
$A^{\prime}\approx0.2$. Note that a system aged up to time $t_{{\rm w}}$
has a ``plateau'' of inactivity for lag times up to $\tau\sim t_{{\rm w}}$.
These plateaus, often associated with the ``caging'' of particles
\cite{Weeks00}, are easily removed with $t/t_{{\rm w}}$ as independent
variable, see Fig.~\ref{fig:msd}(b), leading to the approximate
scaling behavior $\Delta r^{2}\sim\log(t/t_{{\rm w}})$ and a reasonable
data collapse. The residual curvature is removed altogether in panel
$(c)$ using the same correction as in Fig.~\eqref{fig:sisf}, $\epsilon=0.012$,
but a constant $A'\approx0.2$ that is unrelated to $A$.

\begin{figure}
\includegraphics[width=0.5\textwidth]{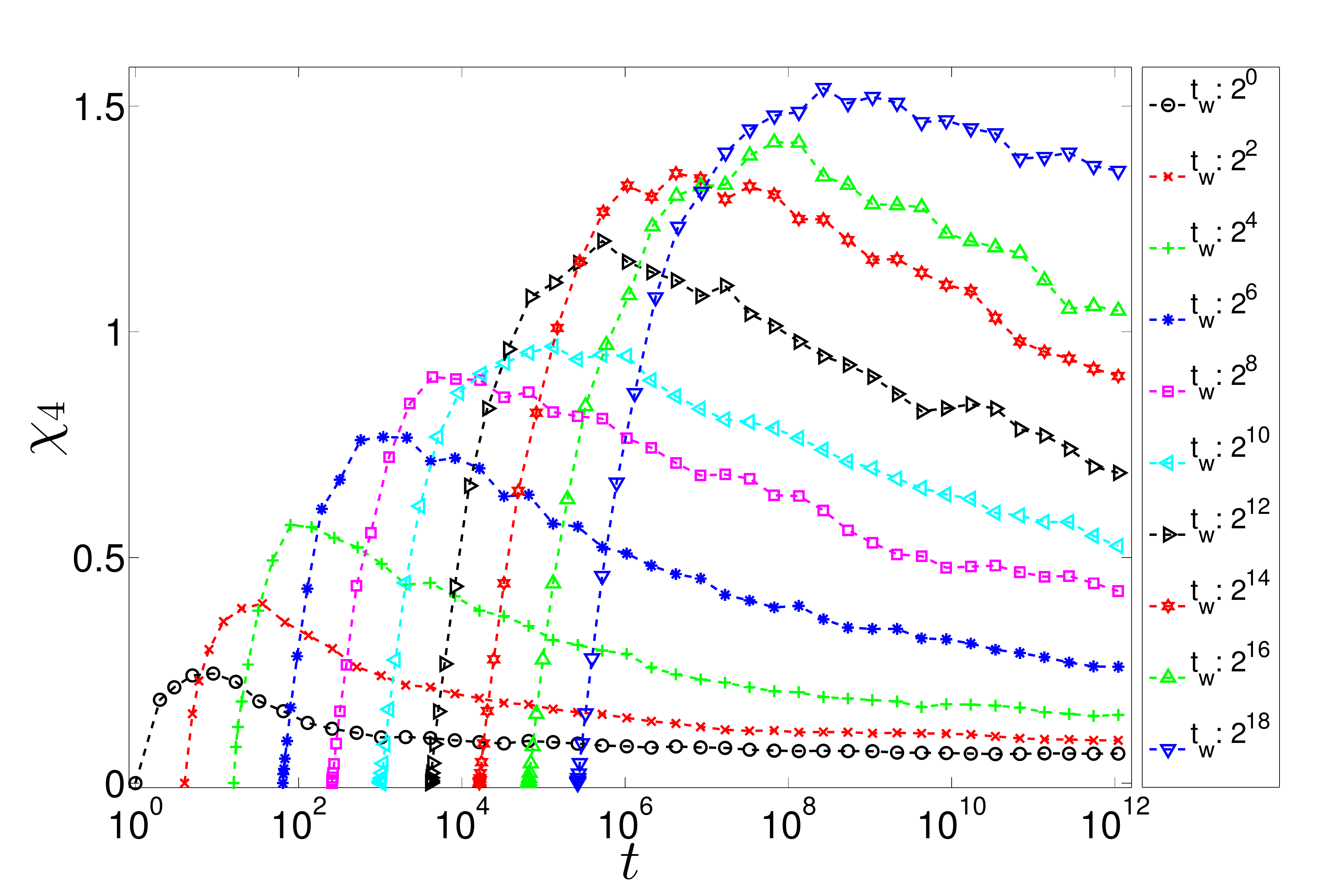} \vspace{-0.5cm}
 \caption{Plot of $\chi_{4}$ as a function of time $t$ for various $t_{{\rm w}}$
at $L=64$. Note that the peak-height increases with $\log t_{{\rm w}}$. }

\label{fig:chi4} \vspace{-0.5cm}
 
\end{figure}

To gauge correlated spatial fluctuation and dynamical heterogeneity,
we consider the 4-point susceptibility \cite{Berthier11} 
\begin{equation}
\chi_{4}(t_{{\rm w}},t)=\langle M(t_{{\rm w}},t)^{2}\rangle-\langle M(t_{{\rm w}},t)\rangle^{2},\label{eq:chi4}
\end{equation}
with mobility measure $M(t_{{\rm w}},t)=\sum_{j}c_{j}(t_{{\rm w}},t)$,
$c_{j}(t_{{\rm w}},t)=\exp\left\{ -\lvert\lvert\vec{\textbf{r}}_{j}(t)-\vec{\textbf{r}}_{j}(t_{{\rm w}})\rvert\rvert_{1}\right\} $,
where $\lvert\lvert\cdot\rvert\rvert_{1}$ denotes the Manhattan norm.
Figure \ref{fig:chi4} shows $\chi_{4}$ vs.~$t$ for a range of
waiting times $t_{{\rm w}}$. While the data does not allow a global
collapse, the logarithmic increase in peak-height vs.~$t_{{\rm w}}$
is clearly visible. This is expected since, by model construction,
heterogeneous dynamics is mainly due to the collapse of clusters,
locked in at a typical size $h\sim\log t_{{\rm w}}$, that re-mobilizes
a corresponding number of particles at a time $t_{{\rm peak}}>t_{{\rm w}}$,
which is reflected in the height of the peak. Fig.~\ref{fig:chi4}
is reminiscent of results in Ref.~\cite{Berthier05} for experiments
and simulations, where $\chi_{4}$ is measured for a sequence of equilibrium
states prepared ever closer to jamming, either by increasing density
or decreasing temperature. 

In summary, our  model coarse-grains
away the ``in-cage rattling'' of particles while incorporating time
intermittency and spatial heterogeneity. Its behavior, which qualitatively
accounts for relevant experimental findings, can be described analytically
using the log-Poisson statistics of cluster collapses~\cite{BoSi09}.

Two-point averages have been plotted versus the lag time,
$\tau=t-t_{{\rm w}}$, to adhere to the established usage and, in
insets, versus  the scaled variable $t/t_{{\rm w}}$. The
first choice  lacks a theoretical basis in the
absence of time translational invariance. The second indicates that
the distinction between an early dynamical regime $\tau<t_{{\rm w}}$
and an asymptotic aging regime $\tau>t_{{\rm w}}$ is moot. 

Deviations from $\log(t/t_{{\rm w}})$-scaling are visible at 
 long times in both the MSD and the SISF (or, equivalently,
the persistence). Interestingly, experimental data show a similar
behavior, see Fig.~1 in Ref.~\cite{BoSi09} and Fig.~4 in Ref.~\cite{ElMasri10}.
As shown in the insets, these deviations can be eliminated  by
 a new scaling variable with a small ($\epsilon\approx1$\%)
correction in $\log(t/t_{{\rm w}})$, whose origin is as follows:
Consider first the fraction of persistent particles $p_{n}$ after
$n$ quakes (marked white in Fig.~\ref{fig:Snapshot}), and neglect
that the size of a quake slowly increases with cluster size and
that particle hits are not uniformly spread throughout the system.
That persistence curve $p_{n}$ then decays exponentially with $n$.
Averaging $p_{n}$ over the Poisson distribution of $n$ gets 
\begin{equation}
{\cal P}(t_{{\rm w}},t)=\exp(-A\mu_{{\rm q}}(t_{{\rm w}},t))=\exp(-A\hat{t}),\label{persistency}
\end{equation}
see Eq.~\ref{eq:f_s_approx}, where $A$ is a small constant and
$\mu_{{\rm q}}(t_{{\rm w}},t)\propto\log(t/t_{{\rm w}})$ as the average
number of quakes occurring between $t_{{\rm w}}$ and $t$. In reality,
spatial heterogeneity means that  quakes increasingly hit the \emph{same}
parts of the system, see Fig.~\ref{fig:Snapshot}, and  that
their effect on persistence hence gradually decreases. Heuristically, this
effect is accommodated by correcting the exponent with the $O(\epsilon)$-term
in $\mu_{{\rm q}}$, as in Eq.~\ref{eq:f_s_approx}. More precisely,
since all moments of the quaking process can be expressed in terms
of $\mu_{{\rm q}}$, the correction is the first term of a Taylor
expansion of the actual exponent. Furthermore, the dependence of cluster
size distribution on $t_{{\rm w}}$ leads to a similar dependence
of $A$. The downward curvature of the MSD plotted vs.~$\hat{t}$
is analogously explained, thus, the weak curvature seen in our data
appears to be a direct consequence of spatial heterogeneity. The mobility
correlation function in Eq.~\eqref{eq:chi4} reveals the presence
of a growing length scale in colloidal systems, here, the average
linear cluster size.

Finally, we suggest three measurements to further elucidate 
colloidal dynamics. The first uses the $\chi_{4}$ susceptibility
function, as we presently do, to investigate a  growing correlation
length as a function of the \emph{age}.
The second collects the PDF of fluctuations in particle positions
over short time intervals of length $\Delta t=at_{{\rm w}}$, where
$a$ is a small constant, uniformly covering the longer interval $(t_{{\rm w}},2t_{{\rm w}})$.
If the rate of intermittent quakes decreases as $1/t_{{\rm w}}$,
the $\log$-Poisson statistic in Eq.~(\ref{persistency}) predicts
PDFs which are \emph{independent} of $t_{{\rm w}}$, as shown in Ref.~\cite{Sibani05}
for  spin-glass simulations. Finally, we suggest
that the slight curvature seen in experiments for MSD vs.~$\log(t/t_{{\rm w}})$~\cite{BoSi09}
and in the tail of the SISF~\cite{ElMasri10} reflects 
 spatial heterogeneity, as discussed above. More
detailed experiments might ascertain if the correction producing  our data
collapse  has a similar effect on experimental data.

NB  thanks the Physics Department at Emory University for
its hospitality. The authors are indebted to the V. Kann Rasmussen
Foundation for financial support. SB is further supported by the NSF
through grant DMR-1207431, and thanks SDU for its hospitality.

\bibliography{Boettcher}

\end{document}